\documentclass[12pt]{article}
\usepackage{amsfonts}
\usepackage{bbm}
\usepackage{mathrsfs}
\usepackage{color}
\usepackage{geometry}             % See geometry.pdf to learn the layout options. There are lots.
\usepackage{graphicx}
\usepackage{amssymb}
\usepackage{amsmath}
\usepackage{epstopdf}
\usepackage{cite}

\usepackage[hyperindex=true,
          pdfstartview=FitH,
          bookmarksnumbered=true,
          bookmarksopen=true,
          citecolor=blue,
          linkcolor=blue,
          colorlinks=true,
          pdfborder=001,
          unicode]{hyperref}

\begin{document}

\title{Flat space compressible fluid as holographic dual of black hole
with curved horizon}
\author{Xin Hao, Bin Wu and Liu Zhao\thanks{Correspondence author}\\
School of Physics, Nankai University, Tianjin 300071, China\\
{\em email}: \href{mailto:shanehowe@mail.nankai.edu.cn}
{shanehowe@mail.nankai.edu.cn},
\href{mailto:wubin@mail.nankai.edu.cn}{wubin@mail.nankai.edu.cn} \\
and
\href{mailto:lzhao@nankai.edu.cn}{lzhao@nankai.edu.cn}}
\date{}
\maketitle

\begin{abstract}
We consider the fluid dual of $(d+2)$-dimensional vacuum Einstein equation
either with or without a cosmological constant. The background solutions admit
black hole event horizons and the spatial sections of the horizons are
conformally flat. Therefore, a $d$-dimensional flat Euclidean space
$\mathbb{E}^d$ is contained in the conformal class of the spatial section of
the black hole horizon. A compressible, forced, stationary and viscous fluid
system can be constructed on the product (Newtonian) spacetime
$\mathbb{R}\times\mathbb{E}^d$ as the lowest order fluctuation modes around
such black hole background. This construction provides the first example of
holographic duality which is beyond the class of bulk/boundary correspondence.\end{abstract}

\section{Introduction}

Since 't Hooft \cite{t'Hooft} and Susskind \cite{susskind} proposed the
so-called holographic principle about twenty years ago, the study of holographic
dual of gravitational systems has become one of the major subjects of study in
the area of high energy theories.
Holography is a property which matches one system in the bulk involving gravity
to another system on the boundary without gravity. The most well understood
realization of holographic duality is the AdS/CFT correspondence
\cite{Maldacena:1997re,Gubser:1998bc,Witten:1998qj},
which establishes an equivalence between the superstring theory on
${\rm AdS}_5 \times {\rm S}^5$ and the four-dimensional $\mathcal {N} = 4$
supersymmetric Yang-Mills gauge theory on the boundary of the $\mathrm{AdS}_5$.
However, there are accumulating evidences indicating that holography can be
realized in situations which requires neither $\mathrm{AdS}$ in the bulk
nor $\mathrm{CFT}$ on the boundary. In this respect, the Gravity/Fluid
correspondence plays a very instructive example.

The relationship between gravity and fluid system was first known
through the membrane paradigm, see \cite{Damor} and
\cite{Price,Damour:2008ji,Eling:2009sj}
for recent applications of this approach.
Later on, such relationship is rediscovered as the long wavelength limit
of AdS/CFT correspondence \cite{Policastro:2002se,Bhattacharyya:2008jc,Baier:2007ix,
Haack:2008cp,Bhattacharyya:2008ji,Bhattacharyya:2008kq,Ashok:2013jda}.
In both approaches, the ratio of the viscosity to the entropy density of the
 dual fluid takes a universal value $\frac{1}{4\pi}$
\cite{Policastro:2001yc,Kovtun:2003wp}.
Considerable efforts have been made
to clarify that the results from the membrane paradigm
and from the long wavelength limit of AdS/CFT are related by
RG flow \cite{Kovtun:2003wp,Son:2007vk,Iqbal:2008by,Bredberg:2010ky}.

Renewed interests in Gravity/Fluid correspondence arise following the
work \cite{strominger1}, in which a fluid dual on a finite
cutoff in a Rindler background has been constructed. This is the first
successful attempt in constructing Gravity/Fluid correspondence
beyond the framework of AdS/CFT. Subsequent works revealed that similar
construction also works in black hole backgrounds in Einstein gravity
\cite{Compere,cai1,Eling,Compere:2012mt} as well as in
higher curvature gravity \cite{Eling:2011cl,Bai:2012ci,Zou:2013ix},
and the viscosity of the dual fluid is calculated in
\cite{Banerjee:2012iz,Hu:2013dza}.
Numerous examples of this correspondence was studied extensively
\cite{Niu:2011gu,Cai:2012mg,Cai,Nakayama:2011bu,1204.2029}.

Besides the membrane paradigm and the AdS/CFT approach, the Gravity/Fluid
correspondence can be realized either using a boost-rescaling technique
\cite{Compere,cai1,Eling,Compere:2012mt,Eling:2011cl,Bai:2012ci,Zou:2013ix,
Hu:2013dza,Niu:2011gu,Cai:2012mg,Cai,Nakayama:2011bu,1204.2029}
or by the introduction of Petrov I boundary conditions
\cite{Strominger2,Ling:2013kua,Ying2,WB1,WB2,Wu:2013kqa,Cai:2014sua,Cai:2014ywa,
xiaoning,Huang:2011kj}, which is mathematically much simpler.
In all known example cases, there are two remarkable features (or drawbacks)
in the Gravity/Fluid correspondence. Firstly, just like in any other
realizations of holographic duality, the boundary (or holographic screen) must
be chosen such that it is an equipotential hypersurface in the bulk spacetime.
This requires, in particular, that if one starts from a spherically symmetric
solution of the gravitational field equations, the final fluid dual must also
live in a spacetime with spherical spatial sections. Thus, if one would like to
understand flat space fluid mechanics from a gravitational perspective,
then the only
choice seems to be starting from an AdS bulk and choose a solution with flat
horizon. Secondly, the dual fluid is always incompressible due to the fact
that to the lowest nontrivial order, the conservation of the Brown-York tensor
on the boundary implies a divergence-free condition of the velocity field of the
dual fluid.

In this work, we are aimed to realize a duality relationship between gravity
solutions with spatially non-flat horizons and a compressible fluid
system living in a flat, Newtonian spacetime with one less dimensions.
Evidently, a flat Newtonian spacetime
cannot be realized as an equipotential hypersurface in the bulk spacetime
unless the gravity solution is plane symmetric. In fact, the flat Newtonian
space may not be a subspace of the bulk spacetime at all. Therefore, the duality
relationship as described above is highly nontrivial not only because it
evades the two drawbacks of general Gravity/Fluid correspondence, but also
because it provides a remarkable example of holographic duality which is beyond
the class of bulk/boundary dualities.

\section{Static black holes in $(d+2)$-dimensions}

Let us start by introducing a particular class of static vacuum solutions to the
Einstein equation
\begin{align}
G_{\mu \nu} = - \Lambda g_{\mu \nu} \label{Ein}
\end{align}
in $(d+2)$-dimensions, where $G_{\mu \nu}=R_{\mu\nu}-\frac{1}{2}g_{\mu\nu}R$ is
the bulk Einstein tensor, $\Lambda$ denotes a possible cosmological constant
which may be positive, zero or negative.

Under the coordinates $x^\mu=(t,r,x^i)$ $(i=1,\cdots,d)$, the static solution to
the vacuum Einstein equation can be described by the line element
\begin{align}
\mathrm{d}s^2_{d+2} = - f(r) \mathrm{d}t^2 + \frac{\mathrm{d}r^2}{f(r)}
+ r^2 e^{\Phi (x^i)}
\delta_{ij} \mathrm{d}x^i \mathrm{d}x^j
\quad , \quad
f(r) = \kappa - \frac{\omega}{r^{d-1}} - \frac{2\Lambda r^2}{d(d+1)} ,  \label{bm1}
\end{align}
where $\kappa=1$ if $\Lambda\geq 0$ and $\kappa=-1, 0, 1$ if $\Lambda<0$.
To ensure that \eqref{bm1} is a solution to the vacuum Einstein equation
\eqref{Ein}, the function $\Phi(x^i)$ must obey a set of complicated
differential equations,
\begin{align}
& \delta^{jk} \partial_j \partial_k \Phi
+ \frac{d-2}{2}\bigg( 2 \partial_i^2 \Phi
+ \delta^{jk} \partial_j \Phi \partial_k \Phi
- (\partial_i \Phi)^2  \bigg) + 2 \kappa (d-1)e^\Phi=0,\nonumber
\\
&\qquad \qquad \qquad (\mbox{no summations over fixed } i=1,\cdots,d)
\label{GLveq} \\
& (d-2) \bigg(\partial_i \partial_j \Phi
- \frac{1}{2} \partial_i \Phi \partial_j \Phi\bigg)=0 \qquad (i \neq j).
\label{GLveq2}
\end{align}
For generic $d$, explicit solution for the function $\Phi(x^i)$ is guaranteed
to exist because the well-known Schwarzschild-Tangherlini-(A)dS solution to the
vacuum Einstein equation can be written in the form \eqref{bm1}.
In particular, when $d=2$, the eqs. \eqref{GLveq}-\eqref{GLveq2} degenerate
into the Euclideanized Liouville equation (or Laplacian equation
if $\kappa=0$) \cite{Moskalets:2014hoa}
\begin{align}
(\partial_1^{\ 2} + \partial_2^{\ 2}) \Phi + 2 \kappa e^\Phi = 0.   \label{Lv}
\end{align}

It is evident that the line element \eqref{bm1} possesses a black hole event
horizon provided $\omega\neq 0$. The event horizon is located at one of
the zeros $r=r_h$ of the metric function $f(r)$, which is the largest (if
$\Lambda$ is non-positive) or the second largest (if $\Lambda$ is positive) root
of $f(r)$. The spatial section of the horizon surface has a conformally
flat metric
\[
\mathrm{d}s_d^2 = r_h^2 e^{\Phi (x^i)}
\delta_{ij} \mathrm{d}x^i \mathrm{d}x^j,
\]
which contains the $d$-dimensional flat Euclidean space $\mathbb{E}^d$
with metric $\delta_{ij}$ in its conformal class.

In what follows, it is desirable to rewrite the line element \eqref{bm1} in the
Eddington-Finkelstein coordinates $(u,r,x^i)$ with the lightlike coordinate $u$
defined by
\begin{align}
u=t+\int_0^r\frac{\mathrm{d}r'}{f(r')}.
\end{align}
Doing so, the line element \eqref{bm1} can be rearranged into the following form
\begin{align}
\mathrm{d}s^2_{d+2}=g_{\mu\nu} \mathrm{d}x^\mu \mathrm{d}x^\nu
=-f(r) du^2 + 2du dr + r^2 e^{\Phi (x^i)} \delta_{ij} \mathrm{d}x^i \mathrm{d}x^j.   \label{bm}
\end{align}

\section{Hypersurface and Brown-York tensor}

Now consider a $(d+1)$-dimensional timelike hypersurface $\Sigma_c$
located at $r=r_c$. The geometry of this embedding hypersurface is best
characterized by its first and second fundamental forms. The first fundamental
form is provided by the restriction of the bulk line element on the
hypersurface, i.e.
\begin{align}
 ds^2_{d+1}=\gamma_{ab}\mathrm{d}x^a\mathrm{d}x^b
 =& -f(r_c)\mathrm{d}u^2 + r_c^2 e^\Phi \delta_{ij}
 \mathrm{d}x^i \mathrm{d}x^j       \nonumber
 \\=& -(\mathrm{d}x^0)^2
 + r_c^2 e^\Phi \delta_{ij} \mathrm{d}x^i \mathrm{d}x^j     \nonumber
\\=& -\frac{1}{\lambda^2} \mathrm{d}\tau^2
+ r_c^2 e^\Phi \delta_{ij} \mathrm{d}x^i \mathrm{d}x^j, \label{dsd}
\end{align}
where $x^a=(u,x^i)$, $\tau = \lambda x^0 = (\lambda \sqrt{f_c})\, u$  and the
rescaling parameter $\lambda$ is introduced to facilitate the forthcoming
analysis on the non-relativistic
limit\footnote{One can think of $1/\lambda$ as the speed of light,
hence $\lambda\to 0$ corresponds to infinite light speed, i.e. the
non-relativistic limit.}.
Here and below we use the notations $f_c = f(r_c)$,
$f'_c = f'(r)|_{r=r_c}$ etc. The notations $f'_h, f''_h$ will also be used,
which are similar to $f'_c, f''_c$ but with $r_c$ replaced by
$r_h$.

One can easily promote the hypersurface tensor $\gamma_{ab}$ to a bulk
tensor $\gamma_{\mu\nu}$ by adding a raw and a column in the $r$-direction
which are both full of zeros.
The second fundamental form is the extrinsic curvature of the
hypersurface, which is defined as
\begin{align}
K_{\mu\nu} = \frac{1}{2} \mathscr{L}_n \gamma_{\mu\nu}, \label{sec}
\end{align}
where
\[
n^\mu = \bigg(\frac{1}{\sqrt{f}},\sqrt{f},0,\cdots,0\bigg)
\]
is a unit vector field which is normal to $\Sigma_c$ at $r=r_c$
and is written in the coordinate $(u,r,x^i)$.

The projection of the bulk Einstein equation gives rise to the
so-called momentum and Hamiltonian constraints,
\begin{align}
&(G_{\mu \nu} + \Lambda g_{\mu \nu})
\gamma^\mu_{\ b} n^\nu |_{\Sigma_c} = 0,  \nonumber\\
&(G_{\mu \nu} + \Lambda g_{\mu \nu})
n^\mu n^\nu |_{\Sigma_c} = 0.           \nonumber
\end{align}
In terms of the two fundamental forms introduced above, these can be
reformulated in the form
\begin{align}
&D_a (K^a_{\ b} - \gamma^a_{\ b} K) = 0  \label{momentum c},\\
&\hat{R} + K^{ab} K_{ab} - K^2 = 2 \Lambda,
\end{align}
where $\hat{R}$ is the Ricci scalar of $\Sigma_c$, $D_a$ is the covariant
derivative that is compatible with $\gamma_{ab}$. Using the definition \cite{Brown:1992br}
\begin{align}
t_{ab}=\gamma_{ab}K-K_{ab}     \label{BYT}
\end{align}
for the Brown-York stress energy tensor $t_{ab}$ on hypersurface,
the momentum constraints \eqref{momentum c} becomes that of the covariant
divergence free condition
\begin{align}
D_a t^a_{\ b} = 0       \label{momc}
\end{align}
for $t_{ab}$, and the Hamiltonian constraint becomes
\begin{align}
\hat{R} + t^a_{\ b} t^b_{\ a} - \frac{t^2}{d}
= 2 \Lambda.    \label{hamc}
\end{align} 
All these are the standard material for the construction of
Gravity/Fluid correspondence.

Before moving on to the construction of dual fluid, let us mention that
the lightlike coordinate $u$ (and hence $\tau$) in the bulk spacetime
becomes naturally timelike on the hypersurface $\Sigma_c$. The time
evolution of the dual fluid will be defined with respect to this coordinate.
However, the spacetime in which the dual fluid lives will not be
$\Sigma_c$ (which is in general curved) but rather the product space
$\mathbb{R}\times\mathbb{E}^d$ in which the first factor represents the time
direction and $\mathbb{E}^d$ is the $d$-dimensional Euclidean space which
lies in the conformal class of the spatial section of $\Sigma_c$. In other
words, the dual fluid will be living in a $(d+1)$-dimensional Newtonian
spacetime.

\section{Petrov I boundary condition}

As usual in Gravity/Fluid correspondence, we introduce the Petrov I
\cite{Coley:2004jv,Coley:2007tp} boundary condition on $\Sigma_c$, i.e.
\begin{align}
C_{(l)i(l)j}=l^\mu(m_i )^\nu l^\rho(m_j)^\sigma C_{\mu\nu\rho\sigma}=0,
\end{align}
where
\begin{align}
&  l^2=k^2=0,\ ,(k,l)=1,\ ,(l,m_{i})=(k,m_{i})=0,\ ,(m_{i},m_{j})=\delta_{ij}
\end{align}
are a set of Newman-Penrose basis vector fields, and $C_{\mu\nu\rho\sigma}$ is
the bulk Weyl tensor. We choose the basis vector fields to be
\begin{align}
&l^\mu
=\frac{1}{\sqrt{2}}\bigg(\frac{1}{\sqrt f}
(\partial_u)^\mu-n^\mu\bigg)
=\frac{1}{\sqrt{2}}\big((\partial_0)^\mu-n^\mu\big),\nonumber\\ 
&k^\mu =\frac{1}{\sqrt{2}}\bigg(\frac{1}{\sqrt f}
(\partial_u)^\mu+n^\mu\bigg)
=\frac{1}{\sqrt{2}}\big((\partial_0)^\mu+n^\mu\big),\nonumber\\
&(m_i)^\mu = r^{-1} e^{-\frac{1}{2}\Phi} (\partial_i)^\mu.
\end{align}
then the boundary condition becomes
\begin{align}
C_{0i0j}+C_{0ij(n)}+C_{0ji(n)}+C_{i(n)j(n)}=0,       \label{pjbc}
\end{align}
where $C_{abcd}$, $C_{abc(n)}$, $C_{a(n)b(n)}$ are projections of the bulk Weyl
tensor
\begin{align}
&C_{abcd}
= \gamma^\mu_{\ a} \gamma^\nu_{\ b} \gamma^\sigma_{\ c} \gamma^\rho_{\ d}
C_{\mu\nu\sigma\rho},         \nonumber
\\& C_{abc(n)} =
\gamma^\mu_{\ a} \gamma^\nu_{\ b} \gamma^\sigma_{\ c} n^\rho
C_{\mu\nu\sigma\rho},      \nonumber
\\& C_{a(n)b(n)}
= \gamma^\mu_{\ a} n^\nu \gamma^\sigma_{\ c} n^\rho C_{\mu\nu\sigma\rho} ,
\nonumber
\end{align}
and all these can be expressed in terms of the the fundamental forms of
$\Sigma_c$:
\begin{align}
&C_{abcd}
= \hat{R}_{abcd} + K_{ad} K_{bc} - K_{ac} K_{bd}
-\frac{4\Lambda}{d(d+1)} \gamma_{a[c} \gamma_{d]b},        \nonumber
\\& C_{abc(n)} = D_a K_{bc} - D_b K_{ac},            \nonumber
\\& C_{a(n)b(n)}
= - \hat{R}_{ab} + K K_{ab} - K_{ac} K^{c}_{\ b}
+ \frac{2 \Lambda}{(d+1)} \gamma_{ab}.            \label{pjwt}
\end{align}
Here, of course, $\hat{R}_{abcd}$ and $\hat R_{ab}$ are the Riemann and Ricci
tensors of $\Sigma_c$.

Then we rewrite the Brown-York tensor (\ref{BYT}) in components
\begin{align}
&K^\tau_{\ \tau}=\frac{t}{d}-t^\tau_{\ \tau}, \qquad
K^\tau_{\ i}=-t^\tau_{\ i},     \nonumber
\\
&K^i_{\ j}=\frac{t}{d}\delta^i_{\ j}-t^i_{\ j},\qquad  K=\frac{t}{d}.
\nonumber
\end{align}
Inserting these relations as well as (\ref{pjwt}) into (\ref{pjbc}), the
boundary conditions finally become
\begin{align}
& \frac{2}{\lambda^2} t^\tau_{\ i} t^\tau_{\ j}
+ \frac{t^2}{d^2} \gamma_{ij} + \frac{2 \Lambda}{d} \gamma_{ij}
- (t^\tau_{\ \tau} - 2 \lambda D_\tau)
\bigg(\frac{t}{d} \gamma_{ij}-t_{ij}\bigg)  \nonumber
\\& \quad- \frac{2}{\lambda} D_{(i} t^\tau_{\ j)}
- t_{ik} t^k_{\ j}
- \hat{R}_{ij}=0,    \label{BYBC}
\end{align}
where the explicit appearance of the parameter $\lambda$ comes from the
rescaling of the coordinate $x^0 \to \tau/\lambda$ which is made in
\eqref{dsd}.

\section{Near horizon and non-relativistic limit}

Now let us place the hypersurface $\Sigma_c$ very close to the black hole event
horizon at $r=r_h$. This means that $r_c-r_h$ is a small parameter, and
we take this parameter to be $r_c-r_h=\alpha^2\lambda^2$, where $\lambda$
is the same rescaling parameter appeared in \eqref{dsd} and $\alpha$ is a finite
constant which must be present to balance the dimensionality.
Note that such an identification implies that the near horizon limit
$\lambda\to 0$ is simultaneously the non-relativistic limit.
The near horizon
nature of $\Sigma_c$ allows us to expand $f_c$ in power series of $\lambda$,
\begin{align}
f(r_c)&=f'(r_h)(r_c-r_h)+\frac{1}{2}f''(r_h)(r_c-r_h)^2+\cdots \nonumber\\
& = f'_h\cdot(\alpha^2\lambda^2) + \frac{1}{2}f''_h \cdot(\alpha^2\lambda^2)^2
+\cdots, \label{fexp}
\end{align}
which is crucial in the following constructions.

To realize the fluid dual of the gravitational theory, it is insufficient to
consider only the background metric. Rather, it is necessary to consider
small fluctuations around the background solution. So, on the hypersurface
$\Sigma_c$, the metric can be expanded in power series in $\lambda$,
\begin{align}
\gamma_{ab} = \gamma_{ab}^{(B)}+\sum_{n=1}^\infty \gamma_{ab}^{(n)}\lambda^n,
\label{gamfluc}
\end{align}
where $\gamma_{ab}^{(B)}$ represents the background metric
and $\gamma_{ab}^{(n)}$ are the fluctuation modes.
Consequently, both the Ricci curvature $\hat R_{ab}$ and the Brown-York tensor
$t^a_{\ b}$ will also be subject to fluctuations, i.e.
\begin{align}
&\hat R_{ab} = \hat R_{ab}^{(B)}
+ \sum^\infty_{n=1} \lambda^n \hat R_{ab}^{(n)},  \label{Rfluc}\\
&t^a_{\ b}= t^{a(B)}_{\ b} + \sum^\infty_{n=1} \lambda^n t^{a(n)}_{\ b},
\label{seby}
\end{align}
where the superscripts $(B)$ indicate contributions from the background
geometry.

In the near horizon limit, the background contributions will also
depend on the parameter $\lambda$, thanks to the expansion \eqref{fexp}. So, we
need to evaluate the background values of the Brown-York and the
Ricci tensors on $\Sigma_c$ and expand the results near the horizon
$r=r_h$. By direct calculations, we can get
\begin{align}
& K^{\tau}_{\ \tau} = \frac{f'_c}{2\sqrt{f_c}} \ \
,\ \ K^{\tau}_{\ i}=0   \nonumber
\\& K^i_{\ j} = \frac{\sqrt{f_c}}{r_c} \delta^i_{\ j}\ \ \ ,
\ \ K = \frac{f'_c}{2\sqrt{f_c}}  + \frac{d\sqrt{f_c}}{r_c}.
\end{align}
These in turn lead to the background Brown-York tensor
\begin{align}
& t^{\tau(B)}_{\ \tau} = \frac{d\sqrt{f_c}}{r_c},      \nonumber\\
& t^{\tau(B)}_{\ i} = 0,             \nonumber\\
& t^{i(B)}_{\ j} = \bigg(\frac{ f'_c }{2\sqrt{f_c}}
+ \frac{(d-1)\sqrt{f_c}}{r_c} \bigg)\delta^i_{\ j},    \nonumber\\
& t^{(B)} = \frac{d}{2}\frac{f'_c}{\sqrt{f_c}}
+ d^2 \frac{\sqrt{f_c}}{r_c}, \label{tb}
\end{align}
where $t^{(B)}$ is the trace of $t^{a(B)}_{\ b}$.
Using \eqref{fexp}, we have the following expansions for
$\frac{\sqrt{f_c}}{r_c}$ and $\frac{f'_c}{\sqrt{f_c}}$:
\begin{align}
\frac{\sqrt{f_c}}{r_c} &=  (\alpha \lambda) \frac{\sqrt{f'_h}}{r_h} +
\frac{1}{2} (\alpha \lambda)^3 \frac{\sqrt{f''_h}}{r_h} + \cdots,
\nonumber\\
\frac{f'_c}{\sqrt{f_c}} &= \frac{\sqrt{f'_h}}{\alpha \lambda}
+ \alpha \lambda \frac{f''_h}{\sqrt{f'_h}} + \cdots. \label{fc}
\end{align}
Inserting the expansions in \eqref{fc} into \eqref{tb} and then into
\eqref{seby}, we get
\begin{align}
&t^\tau_{\ \tau}=\frac{d\alpha\lambda\sqrt{f'_h}}{r_h}
+\lambda t^{\tau(1)}_{\ \tau}+\cdots,   \nonumber\\
&t^\tau_{\ i}=0+\lambda t^{\tau(1)}_{\ i}+\cdots,    \nonumber\\
&t^i_{\ j}=\bigg(\frac{1}{2}\frac{\sqrt{f'_h}}{\alpha\lambda}
+\frac{\alpha\lambda f''_h}{2\sqrt{f'_h}}
+\frac{(d-1)\alpha\lambda\sqrt{f'_h}}{r_h}\bigg)\delta^i_{\ j}
+\lambda t^{i(1)}_{\ j}+\cdots,     \nonumber\\
&t=d\bigg(\frac{1}{2}\frac{\sqrt{f'_h}}{\alpha\lambda}
+\frac{\alpha\lambda f''_h}{2\sqrt{f'_h}}
+\frac{d\alpha\lambda\sqrt{f'_h}}{r_h}\bigg)+\lambda t^{(1)}+\cdots.
\label{PBY}
\end{align}
We shall also make use of the $ij$ components of the Ricci
tensor $\hat R_{ab}$ on $\Sigma_c$. By explicit calculations, we find
\begin{align*}
&\hat \Gamma^{\tau(B)}_{ab}=\hat \Gamma^{a(B)}_{\tau b}=0,\\
&\hat \Gamma^{k(B)}_{ij} =\frac{1}{2}\left(\delta^k{}_i\partial_j\Phi
+\delta^k{}_j\partial_i\Phi - \delta_{ij}\partial^k \Phi\right),
\end{align*}
where $\hat \Gamma^{c(B)}_{ab}$ are components of the Christoffel connection
under the background geometry of $\Sigma_c$. It is remarkable that the
components of $\hat \Gamma^{c(B)}_{ab}$ are independent of the position of
$\Sigma_c$. Consequently, the background Ricci tensor $\hat R_{ab}^{(B)}$
will also independent of $r_c$. Explicitly, we have
\begin{align}
&\hat{R}_{\tau a}^{(B)}=0, \qquad
\hat{R}_{ij}^{(B)}=\kappa(d-1)e^\Phi \delta_{ij}, \label{backR}
\end{align}
where use have been made of the equations \eqref{GLveq} and \eqref{GLveq2}.
So, in the near horizon limit,  $\hat{R}_{ij}^{(B)}$ will not develop
$\lambda$ dependences. However, since the metric $\gamma_{ab}$ on $\Sigma_c$
may fluctuate due to \eqref{gamfluc}, the fluctuation parts $\hat R_{ab}^{(n)}$
in \eqref{Rfluc} will in general be nonzero. Moreover, due to the fluctuations
of the metric, the covariant derivatives such as $D_j t^\tau{}_k$ will also
receive fluctuating corrections which are at least $\mathcal{O}(\lambda^1)$
because
\begin{align}
\hat \Gamma^{c}_{ab} = \hat \Gamma^{c(B)}_{ab} + \mathcal{O}(\lambda^1).
\label{Gafluc}
\end{align}

Finally, substituting \eqref{PBY} and \eqref{Rfluc} (with \eqref{backR}
inserted) into \eqref{BYBC}, we get in the first
nontrivial order $\lambda^{(0)}$ the following identity,
\begin{align}
\frac{\sqrt{f'_h}}{\alpha} t^{i(1)}_{\ j}
&= 2 \gamma^{ik(0)} t^{\tau(1)}_{\ k} t^{\tau(1)}_{\ j}
- 2 \gamma^{ik(0)} \zeta_{kj}
+ \frac{\sqrt{f'_h}}{d\alpha} t^{(1)} \delta^{i}_{\ j},
\label{PBYBC}
\end{align}
where we have introduced the shorthand notations
\begin{align}
&\gamma^{ik(0)}
= r_h^{-2} e^{-\Phi}\delta^{ik},      \nonumber\\
&\zeta_{kj}
= \partial_{(k} t^{\tau(1)}_{\ j)}
- \partial_{(k} \Phi t^{\tau(1)}_{\ j)}
+ \frac{1}{2} \delta_{kj} \delta^{lm}
\partial_l\Phi t^{\tau(1)}_{\ m},       \label{zeta}
\end{align}
which are, respectively, the inverse of the near horizon background metric on
$\Sigma_c$ and the leading term in $D_{(i} t^\tau_{\ j)}$:
\[
D_{(i} t^\tau_{\ j)} =\lambda \zeta_{ij} + \mathcal{O}(\lambda^2).
\]
Some terms which ought to appear in \eqref{PBYBC} cancels out because
\begin{align*}
\frac{2\Lambda}{d} + \frac{f'_h}{r_h}
- \kappa \frac{d-1}{r^2_h} = 0.
\end{align*}

Besides eq.\eqref{PBYBC}, which is the lowest nontrivial order of the Petrov I
boundary condition \eqref{BYBC}, we also need to consider the fluctuation modes
in the covariant conservation condition \eqref{momc}
and the Hamiltonian constraint \eqref{hamc}. Using
\eqref{PBY} and \eqref{Gafluc}, we can evaluate the $\tau$ component of
\eqref{momc}, which reads
\begin{align}
D_at^a_{\ \tau}
&= \partial_\tau t^\tau_{\ \tau} + \partial_i t^i_{\ \tau}
+\hat \Gamma^i_{ij} t^j_{\ \tau}         \nonumber\\
&= -\frac{1}{\lambda}
\bigg[\gamma^{ij(0)}\bigg(\partial_i
+\frac{d-2}{2}\partial_i\Phi\bigg)t^{\tau(1)}_{\ j}
\bigg]+\mathcal{O}(\lambda^{-1}).
\end{align}
Therefore, at order $\lambda^{-1}$, we get
\begin{align}
\delta^{ij}\bigg(
\partial_i + \frac{d-2}{2} \partial_i\Phi
\bigg)  t^{\tau(1)}_{\ j}=0. \label{by1}
\end{align}
Similarly, we can also evaluate the spatial components of
\eqref{momc}, which yields, at the first nontrivial order
$\mathcal{O}(\lambda^{1})$, the following equation,
\begin{align}
\partial_\tau t^{\tau(1)}_{\ i}
-\frac{1}{2} (t^{(1)}-t^{\tau(1)}_{\ \tau}) \partial_i \Phi
+\bigg(\partial_j + \frac{d}{2}\partial_j \Phi\bigg) t^{j(1)}_{\ i}
= 0. \label{by2}
\end{align}
The first non-vanishing order of the
Hamiltonian constraint is at
$\mathcal{O}(\lambda^{0})$:
\begin{align}
t^{\tau(1)}_{\ \tau} = -2 \gamma^{ij(0)}
t^{\tau(1)}_{\ i} t^{\tau(1)}_{\ j}. \label{hamc1}
\end{align}

In the next section we will show that the equations \eqref{PBYBC}, \eqref{by1},
\eqref{by2} and \eqref{hamc1} give rise to the Navier-Stokes equation of a 
forced, stationary and compressible fluid system.

\section{Dual fluid in flat space}

In this section we study the dual fluid equations that arise from the
fluctuation modes described in the last section. For this purpose,
we need to insert \eqref{PBYBC} into \eqref{by2} and simplify the result.
The term $\partial_jt^{j(1)}_{\ i}$ can be evaluated as follows,
\begin{align}\label{PBYT1}
\partial_j t^{j(1)}_{\ i} =
& \frac{\alpha}{\sqrt{f'_h}} \partial_j
\big( 2 \gamma^{jk(0)} t^{\tau(1)}_{\ k} t^{\tau(1)}_{\ i}\big)
- \frac{\alpha}{\sqrt{f'_h}} \partial_j
\big( 2 \gamma^{jk(0)} \zeta_{ki}\big)
+ \frac{1}{d} \partial_i t^{(1)},
\end{align}
where
\begin{align}\label{PBYT1-1}
&\partial_j\big(2\gamma^{jk(0)}t^{\tau(1)}_{\ k}t^{\tau(1)}_{\ i}\big)
=-d\gamma^{jk(0)}\partial_j\Phi t^{\tau(1)}_{\ k} t^{\tau(1)}_{\ i}
+2\gamma^{jk(0)}t^{\tau(1)}_{\ j}\partial_kt^{\tau(1)}_i,
\end{align}
and
\begin{align}\label{PBYT1-2}
&\quad \partial_j \big( 2 \gamma^{jk(0)} \zeta_{ki}\big)
= \partial_i (\gamma^{jk(0)} \partial_j t^{\tau(1)}_{\ k})
+\gamma^{jk(0)}(\partial_j\partial_k - \partial_j\partial_k \Phi
- 2 \partial_j \Phi\partial_k + \partial_j \Phi \partial_k \Phi)
t^{\tau(1)}_{\ i}.
\end{align}
Inserting \eqref{PBYT1-1} and \eqref{PBYT1-2} into \eqref{PBYT1}, we get
\begin{align}
\partial_jt^{j(1)}_{\ i} = &\frac{\alpha}{\sqrt{f'_h}}
\big(-d\gamma^{jk(0)}
\partial_j\Phi t^{\tau(1)}_{\ k}t^{\tau(1)}_{\ i}
+2\gamma^{jk(0)}t^{\tau}_{\ j}\partial_kt^{\tau(1)}_i        \nonumber
-\partial_i(\gamma^{jk(0)}\partial_jt^{\tau(1)}_{\ k})
\\& -\gamma^{jk(0)}
(\partial_j\partial_k
-\partial_j\partial_k\Phi
-2\partial_j\Phi\partial_k
+\partial_j\Phi\partial_k\Phi)
t^{\tau(1)}_{\ i}\big)+\frac{1}{d}\partial_it^{(1)}. \label{ptij}
\end{align}
Substituting \eqref{ptij} as well as \eqref{PBYBC}, \eqref{zeta}
and \eqref{hamc1} into
\eqref{by2}, we get
\begin{align} \label{NSeq2}
&\partial_\tau t^{\tau(1)}_{\ i}+\frac{1}{d}\partial_i t^{(1)}
+ r^{-2}_h e^{-\Phi} \delta^{jk} \bigg[
2 t^{\tau(1)}_{\ k} \partial_jt^{\tau(1)}_{\ i}
- \partial_j\partial_kt^{\tau(1)}_{\ i}  
- 2 t^{\tau(1)}_{\ j} t^{\tau(1)}_{\ k} \partial_i \Phi
       \nonumber
\\&\quad+
\bigg(\partial_j\partial_k\Phi-\frac{d-4}{2}\partial_j\Phi\partial_k
+\frac{d-2}{2}\partial_j\Phi\partial_k\Phi\bigg)
t^{\tau(1)}_{\ i}                           \nonumber
\\&\quad-\partial_j \Phi \partial_i t^{\tau(1)}_{\ k}
+\frac{d-2}{2}(t^{\tau(1)}_{\ k}\partial_j
-\partial_j\Phi t^{\tau(1)}_{\ k})\partial_i\Phi
\bigg] =0,
\end{align}
where we have chosen $\alpha=\sqrt{f'_h}$ to eliminate the constant
factors such as $\frac{\alpha}{\sqrt{f'_h}}$.

Unlike the usual construction of fluid dual, we would like to interpret
eqs. \eqref{by1} and \eqref{NSeq2} as the continuity and the Navier-Stokes
equations respectively in a flat Euclidean space with spatial coordinates
$x^i$. To achieve this, let us first rewrite \eqref{by1} in the following form:
\begin{align}
\partial^j \bigg( e^{\frac{d-2}{2}\Phi} t^{\tau(1)}_{\ j}\bigg)=0. \label{ctn}
\end{align}
Adopting the following ``holographic dictionary''
\begin{align}
\rho = r^2_h e^{\frac{d}{2}\Phi} , \quad
\mu = e^{\frac{d-2}{2}\Phi} ,\quad
\nu = \frac{\mu}{\rho} = r^{-2}_h e^{-\Phi} ,\quad
\end{align}
and
\begin{align}
\quad t^{\tau(1)}_{\ i} = \frac{v_i}{2\nu}  ,\quad
\quad \frac{t^{(1)}}{d}=\frac{p}{2\mu},
\end{align}
where $\rho, \mu, v_i, p$ are respectively the density, viscosity,
velocity field and the pressure of the dual fluid ($\nu$ is the kinematic
viscosity), then eq. \eqref{ctn}
becomes the continuity equation
\begin{align}
\partial^j (\rho  v_j) = 0 , \label{Cont}
\end{align}
and eq. \eqref{NSeq2} becomes the standard Navier-Stokes equation
\begin{align}
\rho (\partial_\tau v_i + v^j \partial_j v_i)
= - \partial_i p + \partial^j d_{ij} + f_i
\label{NV}
\end{align}
for the velocity field of the fluid, where the symmetric traceless tensor
\begin{align}
d_{ij} = \mu \bigg( \partial_j  v_i + \partial_i  v_j
- \frac{2}{d} \delta_{ij} \partial^k v_k \bigg)
\end{align}
represents the deviatoric stress, which depends only on the derivatives of the velocity field and hence vanishes in the hydrostatic equilibrium limit, and
\begin{align}
f_i = \partial^j \Phi \bigg( d_{ij} + \frac{d-2}{2} p \delta_{ij}\bigg)
+ \frac{2}{d} v^j v_j \partial_i \rho
- \frac{2}{d} (v^j \partial_j \rho) v_i   \label{bdyfc}
\end{align}
represents a body force. It is easy to identify the last term in \eqref{bdyfc}
as a linear resistance force, which is proportional to the velocity field
$v_i$ and to the directional gradient of the density of the fluid. The first
two terms in \eqref{bdyfc} look unusual, because the factor $\partial^j\Phi$
actually is proportional to the gradient of the logarithm of the density of the
fluid. Despite the unusual form of the body force, the equations
\eqref{Cont} and \eqref{NV} constitute the complete system of equations
governing the motion of a compressible, forced, stationary and viscous fluid
moving in the $(d+1)$-dimensional Newtonian spacetime
$\mathbb{R}\times\mathbb{E}^d$.

\section{Concluding remarks}

Unlike the ordinary Gravity/Fluid correspondence in which the dual fluid
always lives on an equipotential hypersurface (usually taken to be a near
horizon hypersurface) and is always incompressible, we have constructed
fluid dual in a Newtonian spacetime with one less dimension as compared to the
gravity system. It looks striking to realize such kind of a holographic dual,
because the dual system does not even live on the boundary of the gravitational
system. Some of the distinguished features of our construction are summarized
below:
\begin{itemize}
\item The holographic screen, if one prefers to speak so, is not (necessarily)
a boundary of the bulk spacetime. The dual system lives in a flat Newtonian
spacetime even if the black hole horizon in the bulk is curved;
\item The dual fluid is compressible but stationary, i.e. the density
distribution does not change with time;
\item The dual system possesses a non-constant viscosity;
\item The dual fluid is subject to both the surface stress and a body force,
even though the gravity side is free of source.
\end{itemize}
Going through the construction process, it is clear that our result depends
heavily on the fact that the ``angular part'' of the black hole solution
possesses a conformally flat geometry. This is the case for all maximally
symmetric solutions of Einstein equation in any dimension as well as all
solutions of Einstein equation in the case of $d=2$ (i.e. four dimensional
spacetime) -- in the latter case, the angular
part is two dimensional and it is known that any two dimensional manifold is
conformally flat. To be more concrete, we would like to present the explicit
value of the conformal factor $e^\Phi$ in the case of arbitrary $d\geq2$ with
$\kappa=1$. In this case, we have
\[
e^{\Phi(x)} = \frac{1}{\big(1+\frac{1}{4}\sum_{i} x_i^2\big)^2}.
\]
For the particular choice $d=2$, $e^\Phi$ is given by the solutions of the
Liouville (or Laplacian) equation \eqref{Lv}, and there are infinitely many
different solutions to such equations.

Clearly, much has been left to do following this work. The first question to be
answered is whether similar construction works in the case of other
background geometries or starting from other (extended) theories of gravity
(ether with or without source fields).
Meanwhile, we have chosen to make a near horizon expansion in the intermediate
steps of the construction. Whether the near horizon condition is absolutely
required is in question, because there are already a number of examples in the
ordinary Gravity/Fluid correspondence in which the holographic screen is
not taken as a near the horizon hypersurface but rather as a finite cutoff
surface \cite{Bredberg:2010ky,cai1,Ling:2013kua}
. If all these proves to be working, then a further step will be asking
whether holographic duality beyond the class of bulk/boundary correspondence
can be worked out in more general settings such as Gravity/Condensed Matter
Theory or Gravity/QCD correspondences etc. We hope we could have more to say
following these lines shortly.

\section*{Acknowledgement}
We would like to thank R.-G Cai, Y. Ling, Y. Tian and X. N. Wu for numerous
conversations about Gravity/Fluid correspondence. Bin Wu is supported by
the Ph.D. Candidate Research Innovation Fund of Nankai University.

\providecommand{\href}[2]{#2}\begingroup%\raggedright
\footnotesize\itemsep=0pt
\providecommand{\eprint}[2][]{\href{http://arxiv.org/abs/#2}{arXiv:#2}}

%\bibliographystyle{utcaps}
%\bibliography{papers1,papers2,books}

\end{document}